# Unexpected sawtooth artifact in beat-to-beat pulse transit time measured from patient monitor data

Yu-Ting Lin, M.D., Ph.D.[1], Yu-Lun Lo, M.D.[2], Chen-Yun Lin, Ph.D.[4], Martin G. Frasch, M.D. Ph.D.[3#], Hau-Tieng Wu. M.D. Ph.D.[4,5,6*]

1. Department of Anesthesiology, Taipei Veteran General Hospital, Taipei, Taiwan.
2. Department of Thoracic Medicine, Chang Gung Memorial Hospital, Chang Gung University, College of Medicine, Taipei, Taiwan
3. Department of Obstetrics and Gynecology and Center on Human Development and Disability (CHDD), University of Washington, Seattle, WA
4. Department of Mathematics, Duke University, Durham, NC
5. Department of Statistical Science, Duke University, Durham, NC
6. Mathematics Division, National Center for Theoretical Sciences, Taipei, Taiwan

* Corresponding Author: Hau-tieng Wu. Address: Department of Mathematics and Department of Statistical Science, Duke University, 120 Science Dr. Durham, NC, 27708, USA. Telephone number: 1-858-257-7842, email: hauwu@math.duke.edu

# Co-corresponding Author: Martin Frasch. Address: Department of Obstetrics and Gynecology, University of Washington, Health Sciences Building, 6th Floor, BB Wing, 1959 NE Pacific St, Seattle, WA 98195, USA. Telephone number: 1-206-616-8305, email: mfrasch@uw.edu

# Abstract

**Object**: It is increasingly popular to collect as much data as possible in the hospital setting from clinical monitors for research purposes. However, in this setup the data calibration issue is often not discussed and, rather, implicitly assumed, while the clinical monitors might not be designed for the data analysis purpose. We hypothesize that this calibration issue for a secondary analysis may become an important source of artifacts in patient monitor data. We test an off-the-shelf integrated photoplethysmography (PPG) and electrocardiogram (ECG) monitoring device for its ability to yield a reliable pulse transit time (PTT) signal.

**Approach**: This is a retrospective clinical study using two databases: one containing 35 subjects who underwent laparoscopic cholecystectomy, another containing 22 subjects who underwent spontaneous breathing test in the intensive care unit. All data sets include recordings of PPG and ECG using a commonly deployed patient monitor. We calculated the PTT signal offline.

**Main Results**: We report a novel constant oscillatory pattern in the PTT signal and identify this pattern as a sawtooth artifact. We apply an approach based on the de-shape method to visualize, quantify and validate this sawtooth artifact.

**Significance**: The PPG and ECG signals not designed for the PTT evaluation may contain unwanted artifacts. The PTT signal should be calibrated before analysis to avoid erroneous interpretation of its physiological meaning.

# Introduction

Calibration is one of the most important initial steps in any signal acquisition and experiment -- the data collection equipment, or the quality of the data, needs to be calibrated before a meaningful data analysis can take place. By calibration, we mean the validity of the signal source and checking if the signal is correctly recorded for the specific purpose. While using clinical monitors as scientific instrument has been questioned [1], in our era of medical big data research, we rely on clinical monitors, such as patient vital signs monitors or Holter ECG, heavier than ever before to collect as much data as possible in the hospital research setting for the data analysis purposes [2,20]. While there has been a lot of discussion about the artifact issues in patient monitoring data [3-6], if and how the data is calibrated is often not discussed and, rather, implicitly assumed when collecting and analyzing data acquired from clinical monitors. Particularly, when multiple time series recorded from an off-the-shelf patient monitor are analyzed in the framework of sensor fusion [7], it is often implicitly assumed that on the device level the relationship between channels, such as synchronization, is not an issue.

The calibration problem becomes more severe when we access the publicly available databases. Usually, less background information is available to the data analysts, which precludes a comprehensive judgement of the data quality. For example, while in the MIMIC III waveform database [20] the inter-waveform alignment problem is mentioned, there is no specific quantification of it but only a description. Without a specific quantification of the underlying problem, the information we can extract from the waveform is limited. For most online available databases, in general it is not consistently known which are suitable for which purposes, since the original clinical data acquisition device may not have been designed for the intended purpose of a secondary analysis and we do not have access to the device hardware or software details [17,18,19].

We hypothesize that this less discussed calibration issue for a secondary analysis will become an important source of artifacts in patient monitor data. To the best of our knowledge, this critical validation step in the work flow of any secondary (or even primary) analysis of data collected from clinical acquisition systems has not been reported, with relevance rising, particularly as more massively and passively collected databases become available.

In this paper, we provide an evidence confirming our hypothesis. We analyzed two databases collected passively during patient care in a hospital environment. We identify a calibration problem and the artifact it produces. Specifically, we focus on the pulse transit time (PTT) signal derived from the electrocardiogram (ECG) and photoplethysmography (PPG). We demonstrate that an artifact in the PTT signal [8] referred to as sawtooth artifact can occur because the marketed patient monitor was not designed and calibrated for this specific purpose in the first place.

## Materials and Methods

## Materials

The data set used in the present manuscript comes from two prospective observational studies.

The first study has been approved by the local institutional ethics review boards (Shin

Kong Wu Ho-Su Memorial Hospital, Taipei, Taiwan; IRB No.: 20160706R). Written informed consent was obtained from each patient. From Dec. 2016 to Oct. 2017, we enrolled 33 patients, ASA I to III, scheduled for laparoscopic cholecystectomy (LC).

Inclusion criteria were patients with acute cholecystitis, chronic cholecystitis or gall stone eligible to undergo LC surgery. All surgeries were carried out by one surgeon to ensure the consistency of surgical procedures across all cases.

Exclusion criteria were major cardiac problems, uncontrolled hypertension, arrhythmia shown in pre-operative ECG, known neurological disease, history of drug abuse and anticipated difficult airways.

The recording lasted on average 26.9 minutes with standard deviation of 5.25 minutes and captured the surgical steps from laparoscopic ports establishment to major part of the gallbladder removal.

The physiological data including PPG and ECG were simultaneously recorded from two Philips IntelliVue MP60 Patient Monitors and one Philips IntelliVue MX800 Patient Monitor. Three machines installed in three different operating rooms were used throughout the study. These patient monitors were serviced by an engineer of the manufacturer on quarterly basis.

The data were collected via data dumping system provided by the third-party software, ixTrend Express ver. 2.1 (ixellence GmbH, Wildau, Germany). The sampling rates of ECG (lead II in EASI mode) and PPG channels were 500 Hz and 125 Hz, respectively.

We refer to this first database as SKWHSMH database.

The second study has been approved by the local institutional ethics review boards

(Chang Gang Memorial Hospital, Taipei, Taiwan; IRB No.:104-6531B). Written informed consent was obtained from each patient. From Nov. 2016 to Feb. 2017, we enrolled 22 intubated patients on mechanical ventilation in the intensive care unit who were ready for weaning.

Inclusion criteria were patients that have been intubated for longer than 24 hours, and ready for weaning. Specifically,

- the patient showed clear improvement of the condition which led to mechanical ventilation;

- acute pulmonary or neuromuscular disease or increased intracranial pressure signs were not present; consciousness and semi-recumbency were required;

- $PaO_2 \geq 60mmHg$ and $FiO_2 \leq 40\%$ with PEEP $\leq$ 8cm $H_2O$, or $PaO_2/FiO_2$ >150 mmHg;

- $PaCO_2$<50mmHg or increasing <10% for patients with chronic $CO_2$ retention;

- Heart rate <140 bpm and the systolic blood pressure of 90-160mmHg;

- no vasopressive or inotropic drugs administered for more than 8 hours;

- no intravenous sedation within the previous 24 hours;

- ability for the patient to cough while being suctioned;

- afebrile with the body temperature less than and equal to 38◦ C; negative cuff

leakage test >110ml or >12%.

Exclusion criteria were

- the presence of a tracheostomy,

- or the patient having been on home ventilation prior to ICU admission,

- or the patient's or family's decision not to re-intubate

- or withdrawal from the care anticipated,

- or planned surgery requiring sedation within the next 48 hours.

All recordings lasted 5 minutes during the spontaneous breathing test. The physiological data including PPG and ECG were simultaneously recorded from several Philips IntelliVue MP60 Patient from different beds. These patient monitors were also serviced by an engineer of the manufacturer on quarterly basis.

The data were collected via data dumping system provided by the third-party software, MediCollector ver. 1.0.46 (MediCollector, USA). The sampling rates of ECG (lead II in EASI mode) and PPG channels were 500 Hz and 125 Hz, respectively.

We refer to this second database as CGMH database.

For the reproducibility purpose, the dumped raw ECG signal and the raw PPG signal, and the derived PTT signal [8] of the two databases are made publicly available in Harvard Dataverse [9].

## Methods

We resampled the PPG to 500 Hz by using linear interpolation method for PTT

calculation. To avoid any possible issue caused by the data dumping system, the quality of the dumped signal was visually compared with the signal displayed on the patient monitor. The entire period of the recorded signal was analyzed as described below.

We followed the method of PTT calculation previously reported [10]. The time point of R-peak was determined from the lead II ECG signal. The time point of pulse wave arrival was determined by the maxima of the first derivation during the ascent of the waveform; that is, the location of the fastest ascending PPG waveform. The time interval between each pair of R-peak time and subsequent pulse wave arrival time was calculated and resampled at 4 Hz by the cubic spline interpolation.

To further quantify how the PTT oscillates, particularly locally, and evaluate if the oscillation is physiological, we applied the recently developed de-shape short-time Fourier transform algorithm (dsSTFT) [11]. Compared with the other time-frequency (TF) representations, dsSTFT provides a nonlinear-type TF representation that shows only the fundamental instantaneous frequency of the oscillatory signal [11]. We chose it since the oscillatory pattern in PTT is non-sinusoidal, and most other TF representations will be complicated by the inevitable multiples. For the reproducibility purpose, the dsSTFT code can be downloaded freely from the public domain [12].

## Results

In both databases, we observed an oscillatory pattern in the PTT signal (Fig 1). Although the PTT demonstrates a change that is reciprocal to the blood pressure on the large scale, it contains a sawtooth oscillation with almost constant periods.

## Findings in the SKWHSMH database.

Out of 33 cases, in 11, 19, 2, and 1 subjects, we took a 2000s, 1500s, 1200s and 500s-period of data, respectively, after the stabilization of PPG and ECG signals, as determined by visual inspection of the waveforms. The unified sizes simplified the observation of the sawtooth artifact. Except for one subject, the selected periods covered the majority of the surgical period in each case.

The PTT signals of all subjects and their associated power spectra are shown in Fig 2. It is clear that in all cases, there is a clear sawtooth oscillation in the PTT signal. In the associated power spectra, dominant peaks around 0.01 Hz, 0.012 Hz and 0.1 Hz are observed. Among 33 cases, there are 10 cases with the dominant frequency at 0.01 Hz, 15 cases with the dominant frequency at 0.012 Hz, and 8 cases with the dominant frequency slightly below 0.1 Hz.

The TF representation of the PTT signal by dsSTFT is shown in Fig 3. In each plot, the x-axis denotes the time in seconds, the y-axis is the frequency in Hertz (Hz), and the intensity of the image means the strength of the oscillation inside the PTT at each time and frequency. From the TF representation, there is a dominant line at 0.01 Hz from the beginning to the end in the demonstrated subject. This indicates that at each moment, the PTT signal shows a regular oscillation at 100 seconds period. Coming back to the time series, we see that the artifact is not only oscillating at 0.01 Hz at each moment, but also with the sawtooth pattern. This artifact is the same as those shown in Fig 1.

This local 0.01 Hz oscillation shows up in 10 subjects and the frequency is fixed and persists throughout the LC procedure and regardless of the surgical and anesthetic manipulations. For the other subjects, a local 0.1 Hz oscillation was observed (not shown). To illustrate this fact for all 35 subjects, the mean TF representations of all 35 subjects are shown in the bottom subplot of Fig 2. Although the whole signal was analyzed, for the visualization purpose, only the first 1,600 seconds are shown here, since 1,600 seconds is the shortest recorded signal across all recordings. It is clear that the only dominant curve left after taking average is again the 0.01 Hz curve. Note that although all subjects received LC, the timestamps of different intrasurgical interventions varied. Even under this heterogeneous situation, the 0.01 Hz oscillation persisted. This indicates that this sawtooth oscillation is common across all subjects, which makes it unlikely to be physiological.

## Findings in the CGMH database.

For the CGMH database, we made similar observations. The PTT signals of all subjects and their associated power spectra are shown in Fig 4. There is a sawtooth oscillation in the PTT signal in all cases. In the associated power spectra, there are 9 cases with the dominant frequency at 0.01 Hz (100 second period), 4 cases with the dominant frequency at 0.0067 Hz (150 seconds period), 5 cases with the dominant frequency 0.0133 Hz (75 seconds period), 2 cases with the dominant frequency at 0.1Hz (10 second period), and 2 cases with other dominant frequencies. Compared to the SKWHSMH database, which contains signals longer than 30 minutes, the signals' lengths in the CGMH database are only 5 minutes long. We thus see a more unclear sawtooth pattern on Fig 4 compared to Fig 2. The TF representation of the PTT signal by dsSTFT is shown in Fig 5. Again, we can see a dominant 0.01 Hz in the averaged TF representation. Note that compared with those in Fig 3, the plot is more blurred. It is because the PTT signal in the CGMH database is much shorter. In a 300 seconds period, the artifact of 100 seconds period only appears three times, which degrades the TF representation quality.

## Discussion

The main finding of our study is the quantitative identification of erroneous information in passively collected data from the hospital environment when performing research on a machine that was not designed for this purpose. Specifically, we provide an evidence of such erroneous information from the PTT signal extracted from a clinically widely used monitoring machine that, albeit enabling PTT analysis, was not designed for this purpose. In addition to visualizing the sawtooth artifact in the PTT signal with traditional means in the time domain at ~100 seconds and the power spectrum analysis in the frequency domain, we further apply a modern time-frequency analysis tool, de-shape STFT, to quantify and confirm the sawtooth artifact presenting as a non-physiological oscillation in the PTT signal. While we could not systematically examine all off-the-shelf monitoring devices on the market, we suspect that a similar issue might exist in other machines, other combinations of different channels or may present in other formats. To the best of our knowledge, this is the first reported quantification of such measurement artifact. This finding leads to the conclusion that it is important for clinical researchers to ensure that the data extrapolated from clinical monitors is physiologically accurate, i.e., calibrated for the research purpose, before making any research conclusion.

PPG is ubiquitous in medicine. This optics-based noninvasive signal acquisition technique provides a continuous and convenient display of arterial pulse in finger or in earlobe, and the reading of oxygenation by pulse oximetry. The display of peripheral pulse also allows the assessment of heart and respiratory rates [13,14]. When combined with other physiological signals, PPG provides an even wider spectrum of applications in healthcare. For example, PPG amplitude combined with pulse rate helps the assessment of surgical stress during anesthesia [15]. The addition of ECG helps the adjustment of a cardiac pacemaker, identification and classification of cardiac arrhythmia [14]. PTT is an important application of the combination of ECG and PPG, which can be used as a surrogate of pulse wave velocity to indirectly measure the blood pressure [10,16]. Since the standard patient monitoring instruments that are commonly used in clinical anesthesia and critical care medicine are equipped with ECG and PPG, it is intuitive to measure the PTT by calculating the data exported from the monitor to identify novel predictive features. The observed sawtooth artifact reported in this paper, if not noticed beforehand, might be over-interpreted with a misleading conclusion.

What is the most likely underlying cause for the observed artifact? According to a private communication with a Philips engineer, the patient monitor was not designed for the PTT analysis. However, since the design details of the patient monitor and third-party software algorithms are not accessible to us, we do not have a concrete answer of how it happens and how to correct it. This kind of "black box" issue has been widely discussed in the past few years [17,18,19]. However, to the best of our knowledge, there is still no satisfactory solution up to now. While finding an optimal solution to the current situation is out of the scope of the current paper, we emphasize that without a proper communication channel between the hardware designer and researchers, it is not possible to understand the source of problems when this kind of artifact is encountered. Thus, for researchers using the bedside patient monitors it is necessary to communicate with the manufacturers to obtain as much information as possible about the device before making a conclusion about the captured signal.

Despite the reported quantification of sawtooth artifacts, and the fact that it has been questioned if it is suitable to use clinical monitors as scientific instruments [1] and the above-mentioned black box issue [17,18,19], we have to face the fact that it is currently a trend to use as much data collected from clinical monitors as possible for the "big data" analyses [20]. To efficiently use the massively collected physiological signals from generic medical equipment, particularly those not collected with the research-grade equipment, we need to confirm if the equipment's design permits the off-label use, and if the physiological signal is suitable for the research purpose. Note that while the sawtooth artifact in the demonstrated databases is relatively easily identified, it is conceivable to encounter other, more difficult-to-track artifacts in other setups. The demonstrated sawtooth artifact might misguide study conclusions, or even the following steps, such as like clinical decision making.

From the signal processing perspective, if the nature or pattern of the artifacts can be identified, we can apply signal analysis tools to remove the artifact, and hence maximize the utilization of available physiological signals [21]. For example, in the current sawtooth artifacts, the manifold learning tool proposed for other medical signal analysis can be applied to remove the artifact [22]. See Fig 6 for an example of the artifact removal based on the manifold learning tool. However, while this post-processing could help us maximize the utilization of currently available data, in its present form it may only be applied in special cases such as the regular sawtooth artifact we report here, but not to other kinds of artifacts. Hence, the reliability of the proposed artifact removal approach needs to be further evaluated before it can be applied widely. Since it does not solve the original artifact problem, we do not extensively discuss it in this paper.

This study has several limitations. First, we only consider the universally available specific models from Philips Intellivue and two databases collected in specific clinical settings. The findings may not be generalizable to other patient monitors and clinical setups. A large-scale study with different patient monitors and clinical setups is needed to evaluate the suitability of available equipment for the PTT studies. Second, since the main message to convey is the unexpected artifact in the PTT signal, we do not cover practical points; for example, what is the potential impact of the existence of the sawtooth artifact on the clinical study, and what is the expected benefit by removing those artifacts in PPT calculation. We leave these topics to a future study.

In conclusion, we demonstrate and quantify an artifact in PTT observed in a patient monitor that was not designed for the purpose of PTT analysis, albeit the required channels were integrated into a single machine. These findings reiterate the importance of ascertaining machine clock synchronization and machine-internal data processing before any meaningful multi-channel time series analyses can ensue. This is especially the case when analyzing signals from medical devices not designed for the intended research purpose. Thus, a calibration is needed to confirm the quality of the signal for the intended purpose. Without a suitable calibration, the collected signal might contain unexpected non-physiological patterns, and the data analysis itself might not be meaningful, or even harmful if misleading conclusions are to be drawn.

## Acknowledgement

MGF and HTW thank Uli Kupferschmitt for the discussion of the issue. HTW thanks Professor Kirk Shelley and Professor Aymen Alian for their valuable comments and suggestions.
## References

1. Feldman JM. Can clinical monitors be used as scientific instruments? Anesthesia and Analgesia. 2016; 103(5), 1071–1072.

2. Raghupathi W, Raghupathi V. Big data analytics in healthcare: promise and potential, Health Inf Sci Syst. 2014; 2(1):3

3. Takla G, Petre JH, Doyle DJ, Horibe M, Gopakumaran, B. The Problem of Artifacts in Patient Monitor Data During Surgery: A Clinical and Methodological Review, Anesthesia and Analgesia. 2006;103(5):1196-1204

4. Nizami S, Green JR, McGregor C. Implementation of artifact detection in critical care: A methodological review, IEEE Reviews in Biomedical Engineering. 2013;6:127-142

5. Hravnak M, Chen L, Dubrawski AB, Eliezer C, Gilles P, Michael R. Real alerts and artifact classification in archived multi-signal vital sign monitoring data: implications for mining big data, J. Clin Monit Comput. 2016; 30(6):875-888

6. Chen L, Dubrawski A, Wang D, et al. Using Supervised Machine Learning to Classify Real Alerts and Artifact in Online Multi-signal Vital Sign Monitoring Data, Crit. Care Med. 2016; 44(7):e456-e463
15

7. Gravina R, Alinia P, Ghasemzadeh H, Fortino G. Multi-sensor fusion in body sensor networks: state-of-the-art and research challenges, Inf. Fusion, 2017; 35:68–80.

8. Mukkamala R, Hahn JO, Inan OT, et al. Towards Ubiquitous Blood Pressure Monitoring via Pulse Transit Time: Theory and Practice, IEEE Trans Biomed Eng. 2015;62(8):1879-1901

9. https://dataverse.harvard.edu/dataset.xhtml?persistentId=doi:10.7910/DVN/OJBZ67

10. Gesche H, Grosskurth D, Küchler G, Patzak A. Continuous blood pressure measurement by using the pulse transit time: comparison to a cuff-based method. European journal of applied physiology. 2012; 112(1):309-315.

11. Lin CY, Su L, Wu HT. Wave-shape function analysis – when cepstrum meets time-frequency analysis, Journal of Fourier Analysis and Applications. 2018; 24(2):451-505

12. https://hautiengwu.wordpress.com/code/

13. Shelley KH. Photoplethysmography: beyond the calculation of arterial oxygen saturation and heart rate. Anesthesia and analgesia. 2007;105(6 Suppl):S31.

14. Tang SC, Huang PW, Hung CS. Identification of Atrial Fibrillation by Quantitative Analyses of Fingertip Photoplethysmogram. Scientific Reports. 2017;7.

15. Huiku M, Uutela K, Van Gils M, et al. Assessment of surgical stress during general anaesthesia. British journal of anaesthesia. 2007;98(4):447-455.

16. Kim SH, Song JG, Park JH, Kim JW, Park YS, Hwang GS. Beat-to-beat tracking of systolic blood pressure using noninvasive pulse transit time during anesthesia induction in hypertensive patients. Anesthesia and analgesia. 2013;116(1):94.

17. Cannesson M, Shafer SL. All boxes are black. Anesthesia and Analgesia. 2016;122(2), 309–317.

18. Ruskin KJ, Shelley KH. Patent medicine and the "black box." Anesthesia and Analgesia. 2005;100(5), 1361–1362.

19. Shelley KH, Barker, SJ. Disclosures, what is necessary and sufficient? Anesthesia and Analgesia. 2016;122(2), 307–308.

20. Johnson AEW, Pollard TJ, Shen L, Lehman L, Feng M, Ghassemi M, Moody B, Szolovits P, Celi LA, Mark RG. MIMIC-III, a freely accessible critical care database. Scientific Data; 2016;3, 160035.

21. Lu Y, Wu HT, Malik J. Maximize information quantity for mobile health – recycling the cardiogenic artifact in impedance plethysmography as an example. Biomedical Signal Processing & Control. 2019; 51, 162-170

22. Su L, Wu HT. Extract fetal ECG from single-lead abdominal ECG by de-shape short time Fourier transform and nonlocal median. Frontiers in Applied Mathematics and Statistics. 2017;3:2.

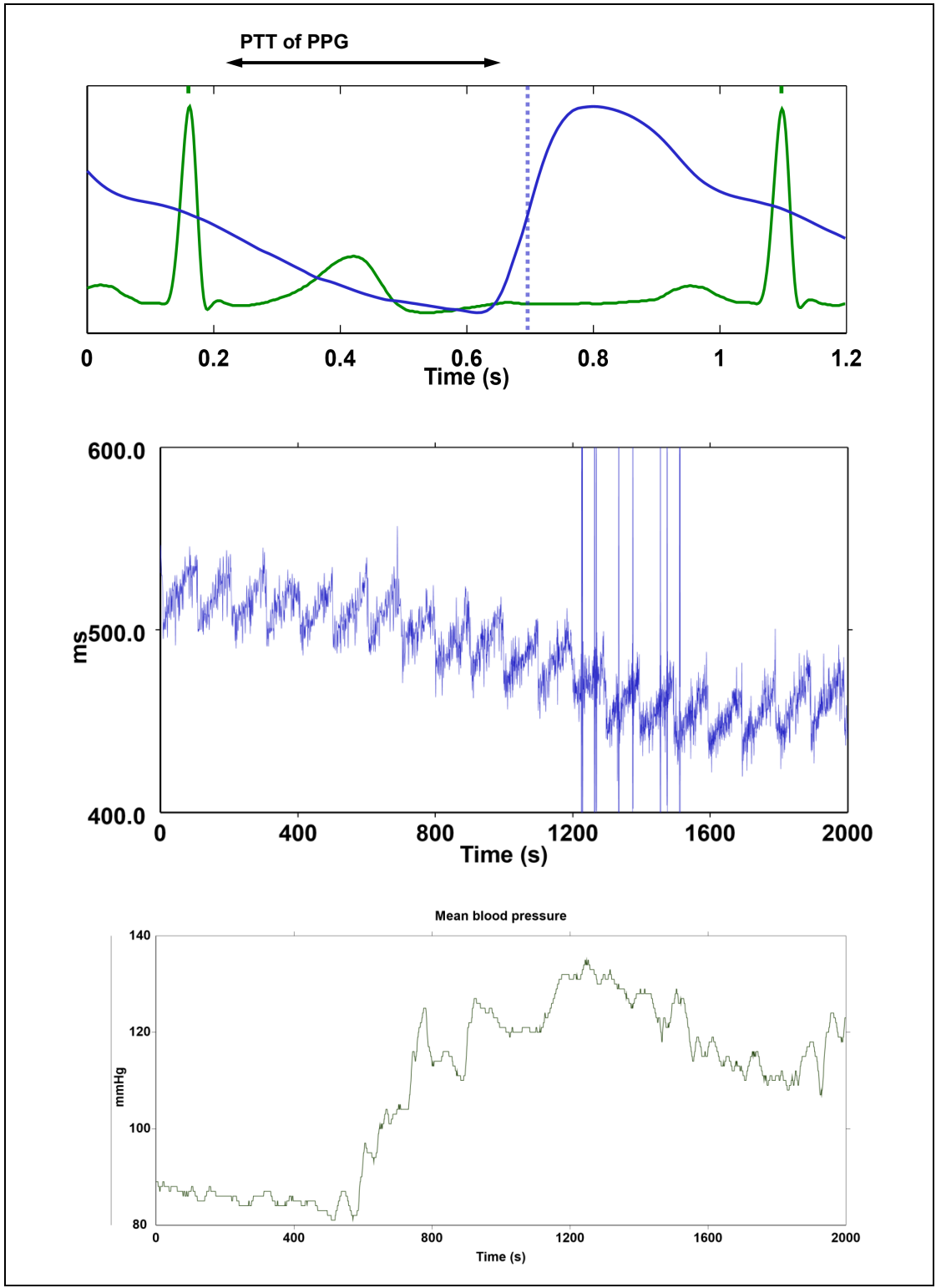

Figure 1. Upper panel: an illustration of the calculation of PTT

Photoplethysmography (PPG). The PPG is shown in blue, and the electrocardiogram

is shown in green. The blue dash line indicates the timestamps of landmarks of PPG

for the calculation of PTT.[10] Middle panel: PTT calculated from PPG. Lower panel: mean blood pressure measured by direct arterial blood pressure waveform. There is a marked sawtooth pattern in the PTT. This sawtooth pattern shows periodicity of 100 s.

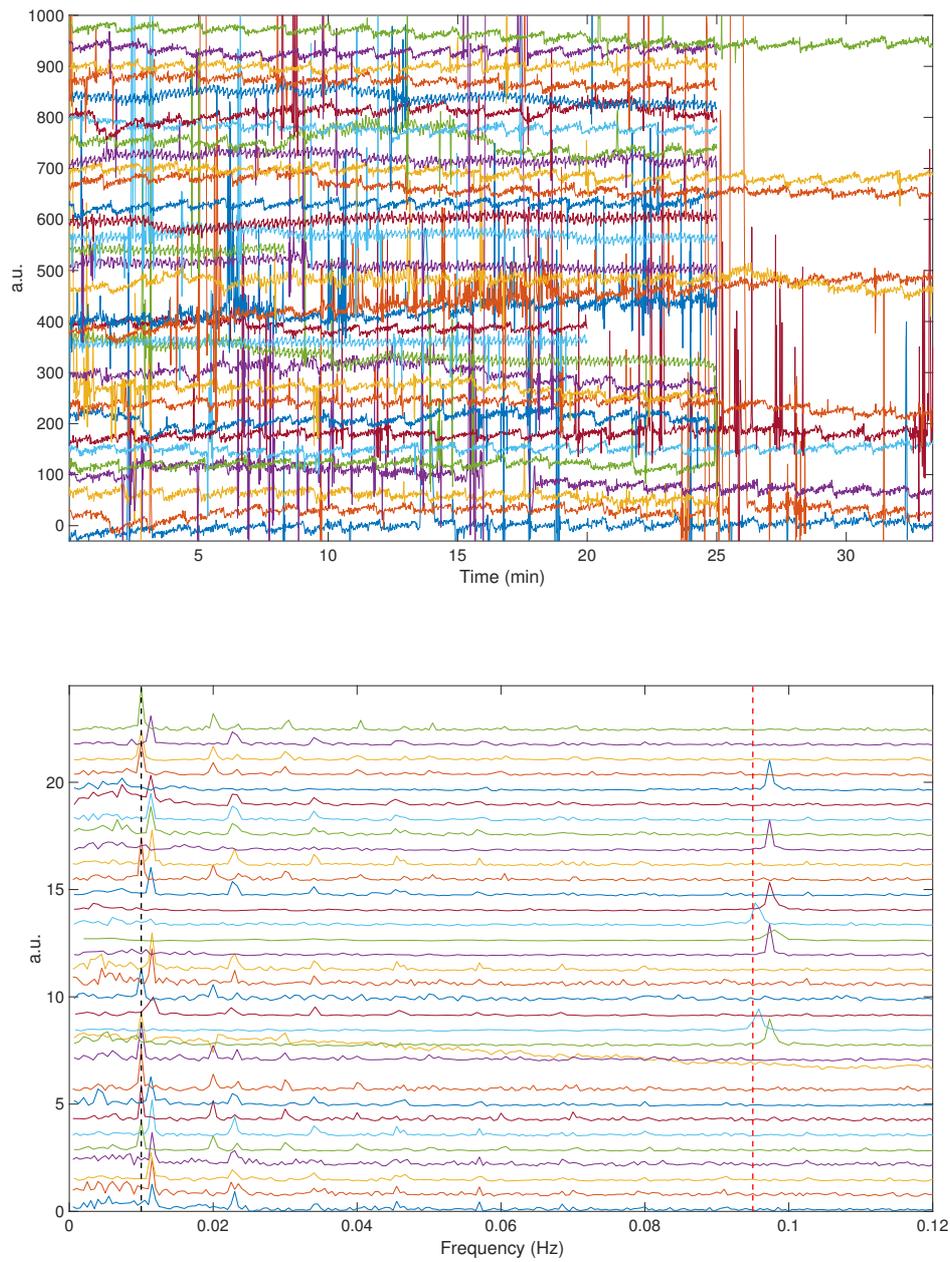

Figure 2. The pulse transit time (PTT) during the entire laparoscopic cholecystectomy procedure of all enrolled subjects in the SKWHSMH database is shown in the top subplot, and their associated power spectra are shown in the bottom subplot. The PTT signals and the power spectra are shifted to enhance visualization. The black dash line and red dash line in the bottom subplot denote the frequencies of 0.01 Hz and 0.095 Hz, respectively. a.u. = arbitrary unit.

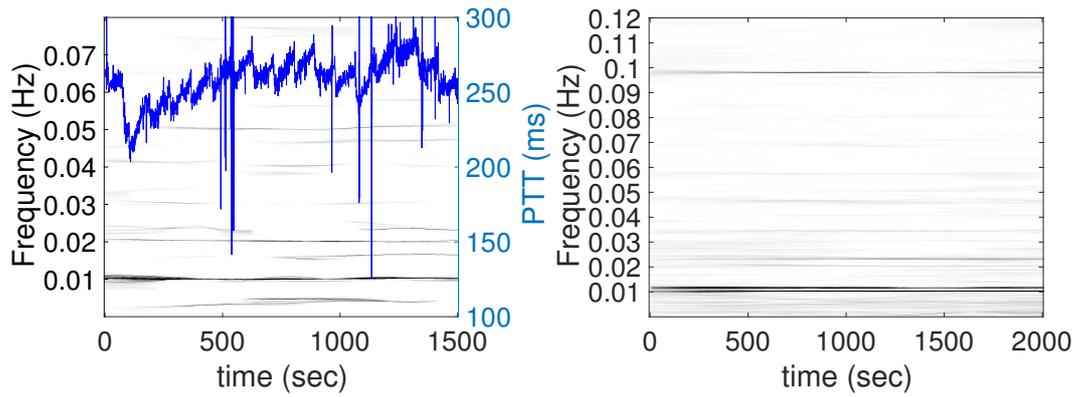

Figure 3. The time-frequency (TF) representation of the PTT (blue curve) determined by the de-shape short-time Fourier transform of subjects from the SKWHSMH database. Left figure shows the PTT from a subject and its TF representation. Right figure shows the average TF representation over 35 subjects. It is clear that there is a dominant line at 0.01 Hz and a line slightly below 0.1 Hz. This indicates a regular oscillation of 0.01 Hz, or a regular oscillation of about 0.1 Hz in most of PTT signals.

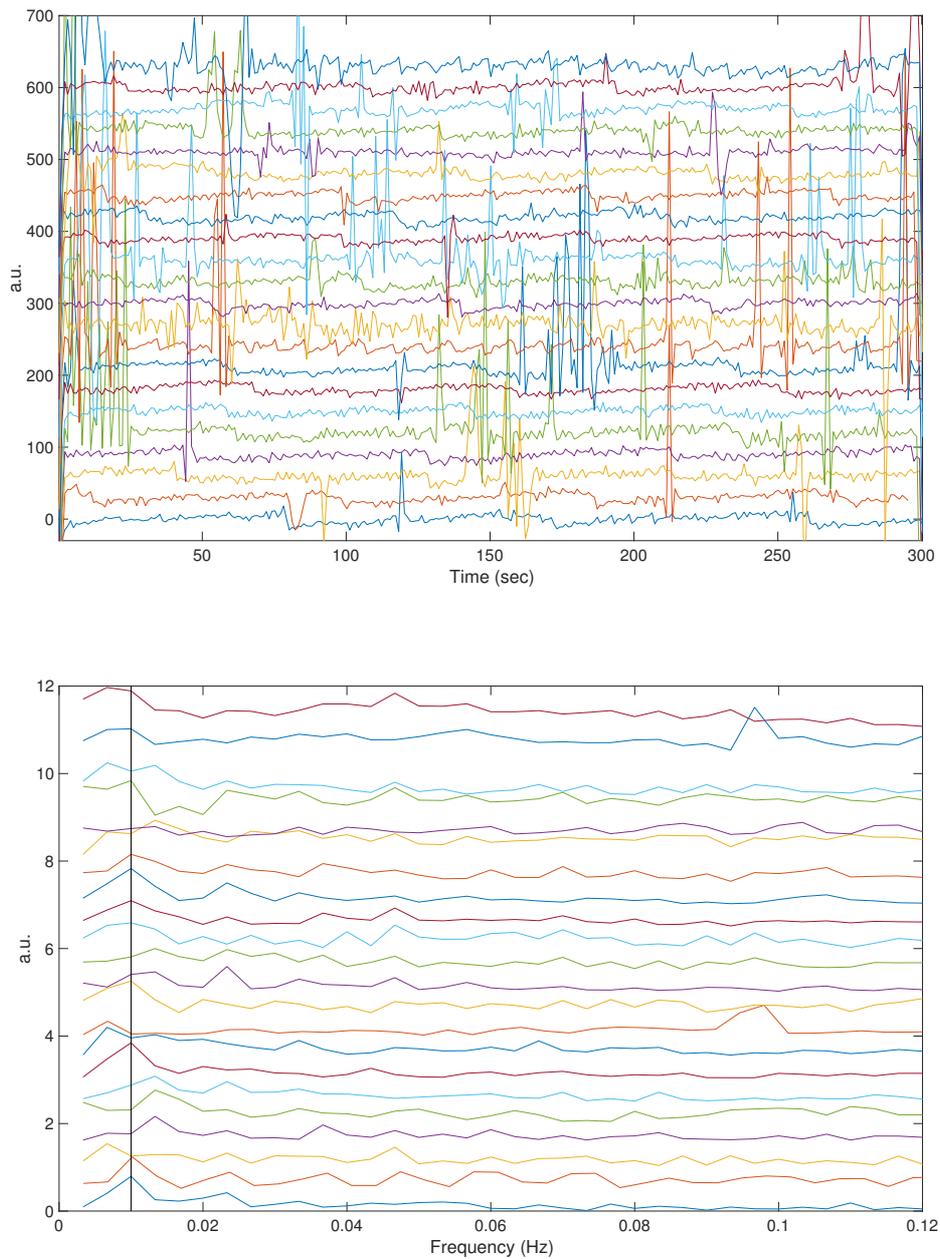

Figure 4. The pulse transit time (PTT) of subjects in the CGMH database are shown in the first subplot, and their associated power spectra are shown in the second subplot. The PTT signals and the power spectra are shifted to enhance visualization. Note that compared with Figure 2, in this database the signal is of only 5 minutes long, so the sawtooth artifact is stretched and there are only few sawtooth cycles. The black line in the second subplot indicates 0.01 Hz. a.u. = arbitrary unit.

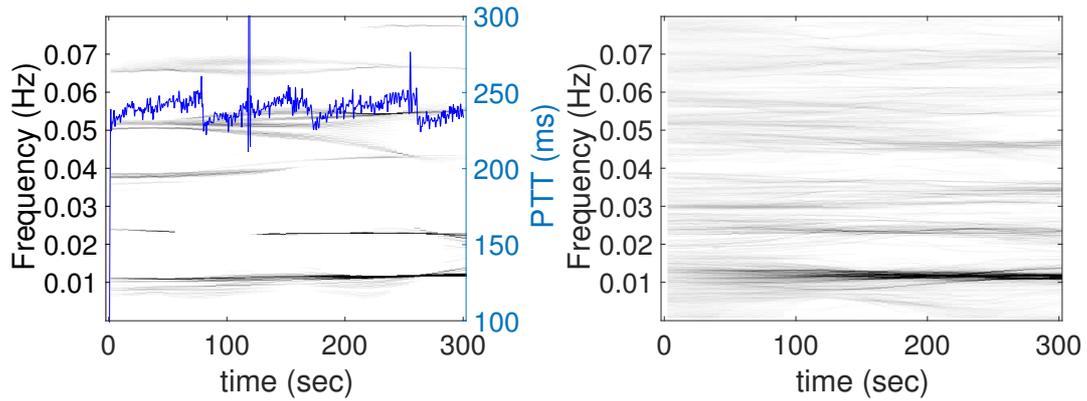

Figure 5. The time-frequency (TF) representation of the PTT (blue curve) determined by the de-shape short-time Fourier transform of subjects from the CGMH database. Left figure shows the PTT from a subject and its TF representation. Right figure shows the average TF representation over 22 subjects. It is clear that there is a dominant line at 0.01Hz. This indicates a regular oscillation of 0.01Hz in most PTT signals.

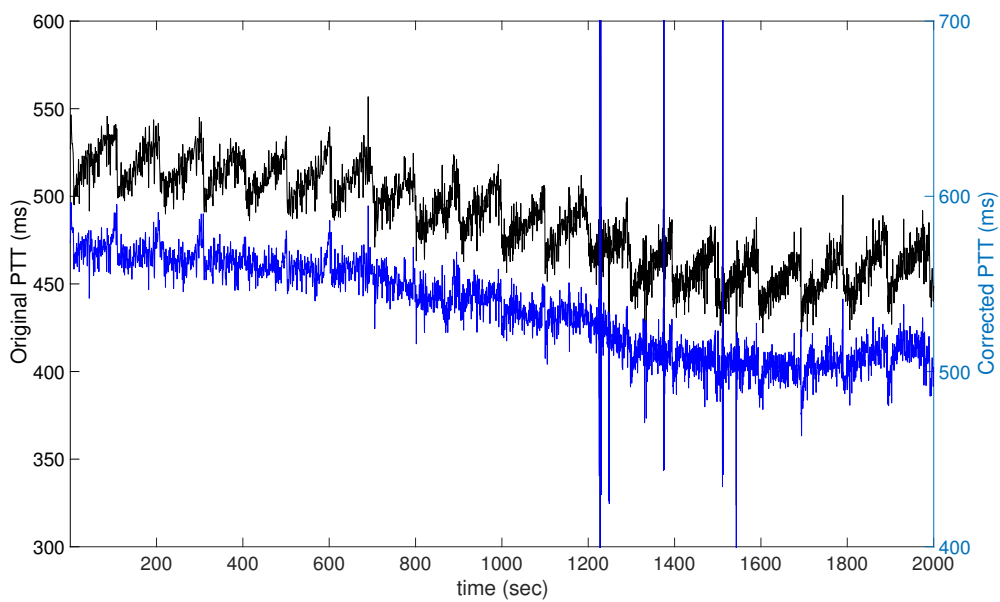

Figure 6. An illustration of removing the sawtooth artifact by the manifold learning tool. The black tracking is the original PTT signal (with the unit shown on the left

ticks), while the blue one is the corrected PTT signal (with the unit shown on the right ticks).